\documentclass[twocolumn,preprintnumbers,amsmath,amssymb]{revtex4}
\usepackage{graphicx}
\usepackage{dcolumn}
\usepackage{bm}

\begin{document}
\title{The Forgotten Life of Mileva Mari{\'c} Einstein}

\author{Pauline Gagnon}

\affiliation{Retired Senior Research Scientist, Indiana University}

\date{February 11, 2020}

\begin{abstract}
Today, February 11, is the International Day of Women and Girls in Science, the perfect day to remember Mileva Mari\'{c}  Einstein, a brilliant but largely unknown scientist. While her husband, Albert Einstein is celebrated as perhaps the best physicist of the century, one question about his career remains: How much did his first wife contribute to his groundbreaking science? While it is not possible to credit her with any specific part of his work, their letters and numerous testimonies presented in books dedicated to her $^{[1-5]}$ provide substantial evidence on how they collaborated from the time they met in 1896 up to their separation in 1914. They depict a couple united by a shared passion for physics, music and for each other. So here is her story.
\\
\end{abstract}

\maketitle

\begin{figure}[t!]
\includegraphics[width=0.5\textwidth]{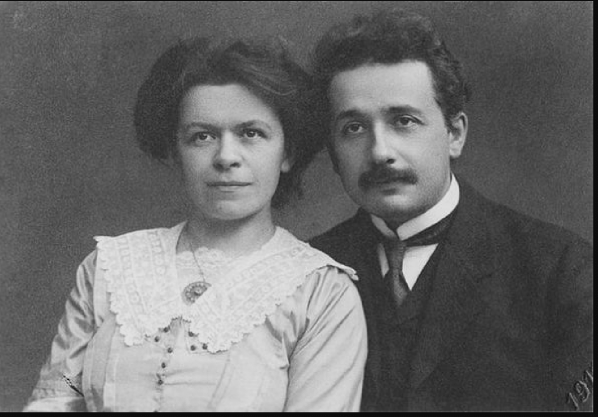}
\caption{Mileva Mari\'{c} Einstein and her husband, Albert Einstein in 1912. Credit: ETH Zurich Archives (CC BY-SA 4.0)}
\end{figure}

Mileva Mari\'{c} was born in 1875 in Titel in Vojvodina, which is now part of Serbia but was then in the Austro-Hungarian Empire. Her parents, Marija Ruzi\'{c} and Milo\v{s} Mari\'{c}, a wealthy and respected member of his community, had two other children: Zorka and Milo\v{s}  Jr. Mileva attended high school the last year girls were admitted under the Austro-Hungarian rules. In 1892, her father obtained the authorisation of the Minister of Education to allow her to attend physics lectures reserved to boys. She completed her high school in Zurich in 1894 and her family then moved to Novi Sad. Mileva's classmates described her as brilliant but reserved. She liked to get to the bottom of things, was perseverant and worked towards her goals.

Albert Einstein was born in Ulm in Germany in 1879 and had one sister Maja. His father, Hermann, was an industrial. His mother, Pauline Koch came from a rich family. Albert was inquisitive, bohemian and rebel. Being undisciplined, he hated the rigour of German schools so he too finished his high school in Switzerland and his family relocated to Milan.

Albert and Mileva were admitted to the physics-mathematics section of the Polytechnic Institute in Zurich (now ETH) in 1896 with three other students: Marcel Grossmann, Louis Kollros and Jakob Ehrat. Albert and Mileva became inseparable, spending countless hours studying together. He attended only a few lectures, preferring to study at home. Mileva was methodical and organized. She helped him channel his energy and guided his studies as we learn from Albert's letters, exchanged between 1899-1903 during school holidays: 43 letters from Albert to Mileva have been preserved but only 10 of hers remain.$^{[5]}$ These letters provide a first-hand account on how they interacted at the time.

In early August 1899, Albert wrote to Mileva: {\it "When I read Helmholtz for the first time, I could not, and still cannot, believe that I was doing so without you sitting next to me. I enjoy working together very much, and find it soothing and less boring."}$^{[5]}$ Then on 28 September 1899, he wrote from Milan: {\it "I am starting to feel uneasy here; the climate does not suit me, and without anything in particular to work on, I tend to brood a lot – in short, I am beginning to feel the absence of your beneficent thumb, under which I am always kept in line."}$^{[5]}$

Mileva boarded in a pension for women where she met her life-long friends Helene Kaufler-Savi\'{c} and Milana Bota. Both spoke of Albert's continuous presence at Mileva's place, where he would come freely to borrow books in her absence. Milan Popovi\'{c}, Helene's grandson, published the letters Mileva exchanged with her throughout her life.$^{[4]}$ 

By the end of their classes in 1900, Mileva and Albert had similar grades (4.7 and 4.6, respectively) except in applied physics where she got the top mark of 5 but he, only 1. She excelled at experimental work while he did not. But at the oral exam, Professor Minkowski gave 11 out of 12 to the four male students but only 5 to Mileva. Only Albert got his degree. 

Meanwhile, Albert's family strongly opposed their relationship. His mother was adamant. {\it "By the time you're 30, she'll be an old witch!"} as Albert reported to Mileva in a letter dated 27 July 1900, as well as: {\it "No decent family will have her."}$^{[5]}$ Mileva was neither Jewish, nor German. She had a limp and was too intellectual in his mother's opinion, not to mention prejudices against foreign people. Moreover, Albert's father insisted his son found work before getting married.
  
On 13 September 1900, Albert wrote to Mileva: {\it "I am also looking forward to working on our new papers. You must now continue with your investigations – how proud I will be to have a little Ph.D. for a sweetheart when I remain a completely ordinary person."}$^{[5]}$ They both came back to Zurich in October 1900 to start their thesis work. The other three students all received assistant positions at the Institute, but Albert did not. He suspected that professor Weber was blocking him. Without a job, he refused to marry her. They made ends meet by giving private lessons and were {\it "living and working as before"}$^{[4]}$ as Mileva wrote to her friend Helene Savi\'{c} in the winter of 1901.

On 13 December 1900, they submitted to the Annalen der Physik a first article on capillarity signed only under Albert's name. Nevertheless, both referred to this article in letters as their common article. Mileva wrote to Helene Savi\'{c} on 20 December 1900. {\it "We also sent a copy to Boltzmann and we would like to know what he thinks of it; I hope he will write to us."}$^{[4]}$ Likewise, Albert wrote to Mileva on 4 April 1901, saying that his friend Michele Besso {\it "went on my behalf to visit his uncle, Prof. Jung, one of the most influential professors in Italy to give him our paper."}$^{[5]}$ 

The decision to publish only under his name seems to have been taken jointly. Why? Radmila Milentijevi\'{c}, an Emeritus history professor at the City College of the City University in New York, published in 2015 Mileva's most comprehensive biography.$^{[1]}$ She suggests that Mileva probably wanted to help Albert make a name for himself, such that he could find a job and marry her. Moreover, Milentijevi\'{c} argues that Mileva had been raised in a very patriarchal family where a woman was expected to do anything to support her husband. Dord Krsti\'{c}, a former physics professor at Ljubljana University, spent 50 years researching Mileva's life. In his well-documented book,$^{[2]}$ he suggests that given the prevalent bias against women at the time, a publication co-signed with a woman might also have carried less weight. 

We will never know. But nobody made it clearer than Albert Einstein himself that they later collaborated on special relativity when he wrote to Mileva on 27 March 1901: {\it "I'll be so happy and proud when we are together and can bring our work on relative motion to a successful conclusion."}$^{[5]}$  

Then Mileva's destiny changed abruptly. She became pregnant after a lovers' escapade in Lake Como. Unemployed, Albert would still not marry her. With this uncertain future, Mileva took her second and last attempt at the oral exam in July 1901. This time, Prof. Weber, whom Albert suspected of blocking his career, failed her. Forced to abandon her studies, she went back to Serbia, but came back briefly to Zurich to try to persuade Albert to marry her. She gave birth to a girl named Liserl in January 1902. No one knows what happened to her. She was probably given to adoption. No birth or death certificates were ever found.

Earlier in December 1901, their classmate Marcel Grossman's father intervened to get Albert a post at the Patent Office in Bern. He started work in June 1902. In October, before dying, his father granted him his permission to marry. Albert and Mileva married on 6 January 1903. Albert worked 8 hours a day, 6 days a week at the Patent Office while Mileva assumed the domestic tasks. In the evenings, they worked together, sometimes late in the night. Both mentioned this to friends, he to Hans Wohlwend, she to Helene Savi\'{c} on 20 March 1903 where she expressed how sorry she was to see Albert working so hard at the office. On 14 May 1904, their son Hans Albert was born.
 
Despite these circumstances, 1905 is known as Albert's "miracle year": he published five articles: one on the photoelectric effect (which led to the 1921 Nobel Prize), two on Brownian motion, one on special relativity and the famous $E = mc^2$. He also commented on 21 scientific papers for a fee and submitted his thesis on the dimensions of molecules. Much later, Albert told R. S. Shankland $^{[6]}$ that relativity had been his life for seven years and the photoelectric effect, for five years. Peter Michelmore, one of his biographers, wrote that after having spent five weeks to complete the article containing the basis of special relativity, Albert {\it "went to bed for two weeks. Mileva checked the article again and again, and then mailed it."}$^{[7]}$  Exhausted, the couple made the first of three visits to Serbia where they met numerous relatives and friends, whose testimonies provide a wealth of information on how Albert and Mileva collaborated.

Mileva's brother, Milo\v{s} Jr, a person known for his integrity, stayed on several occasions with the Einstein family. While studying medicine in Paris, he took a semester off to study in Bern. Krsti\'{c} wrote: {\it "[Milo\v{s}] described how the evenings and at night, when silence fell upon the town, the young married couple would sit together at the table and by the light of a kerosene lantern, they would work together on physics problems. Milo\v{s} Jr. spoke of how they calculated, wrote, read and debated."}$^{[2]}$ Krsti\'{c} heard this directly from relatives of Mileva, Sidonija Gajin and Sofija Gali\'{c} Golubovi\'{c}. 

Zarko Mari\'{c}, a cousin of Mileva's father, lived at the countryside property where the Einstein family stayed during their visit. He told Krsti\'{c} how Mileva calculated, wrote and worked with Albert. The couple often sat in the garden to discuss physics. Harmony and mutual respect prevailed. Gajin and Zarko Mari\'{c} also reported hearing from Mileva's father that during the Einstein's visit to Novi Sad in 1905, Mileva confided to him: {\it "Before our departure, we finished an important scientific work which will make my husband known around the world."}$^{[2]}$ Krsti\'{c} got this same information in 1961 from Mileva's cousin, Sofija Gali\'{c} Golubovi\'{c}, who was present when Mileva said it to her father.

Desanka Trbuhovi\'c-Gjuri\'{c} published Mileva's first biography in Serbian in 1969.$^{[3]}$ It later appeared in German and French. She described how Mileva's brother often hosted gatherings of young intellectuals at his place. During one of these evenings, Albert would have declared: {\it "I need my wife. She solves for me all my mathematical problems,"}$^{[3]}$ something Mileva is said to have confirmed. 

In 1908, the couple constructed with Conrad Habicht an ultra-sensitive voltmeter. Trbuhovi\'c-Gjuri\'{c} attributes this experimental work to Mileva and Conrad, and wrote: {\it "When they were both satisfied, they left to Albert the task of describing the apparatus, since he was a patent expert."}$^{[3]}$ This patent was registered under the name Einstein-Habicht. When Habicht questioned Mileva's choice not to include her name, she replied making a pun in German: {\it "Warum? Wir beide sind nur ein Stein. (Why? The two of us are but one stone)}", meaning, we are one.

The first recognition came in 1908. Albert gave unpaid lectures in Bern, then was offered his first academic position in Zurich in 1909. Mileva was still assisting him. Eight pages of Albert's first lecture notes are in her handwriting. So is a letter drafted in 1910 in reply to Max Planck who had sought Albert's opinion. Both documents are kept in the Albert Einstein Archives (AEA) in Jerusalem. On 3 September 1909, Mileva confided to Helene Savi\'c: {\it "He is now regarded as the best of the German-speaking physicists, and they give him a lot of honours. I am very happy for his success, because he fully deserves it; I only hope and wish that fame does not have a harmful effect on his humanity."}$^{[4]}$ In the Winter 1909-10, she added: {\it "You see, with that kind of fame he does not have much time left for his wife. (...) What is there to say, with notoriety, one gets the pearl, the other the box."}$^{[4]}$

Their second son, Eduard, was born on 28 July 1910. Up to 1911, Albert still sent affectionate postcards to Mileva. But in 1912, he started an affair with his cousin, Elsa L\"owenthal while visiting his mother who had moved to Berlin. They maintained a secret correspondence over two years. Elsa kept 21 of his letters, now in the Collected Papers of Albert Einstein $^{[8]}$. During this period, Albert held various faculty positions first in Prague, back in Zurich and finally in Berlin in 1914 to be closer to Elsa.

This affair caused their marriage's collapse. Mileva moved back to Zurich with her two sons on 29 July 1914. In 1919, she agreed to divorce, with a clause stating that if Albert ever received the Nobel Prize, she would get the money. When she did, she bought three small apartment buildings and lived decently from their income until the economic crash of 1929. After that, her tenants could no longer pay their rent and she soon lost two of those buildings. 

During his childhood, her youngest son Eduard frequently stayed in a sanatorium. He later developed schizophrenia and was eventually internalised. Due to these medical expenses, Mileva struggled financially especially at the end of her life. She survived by giving private lessons and on the alimony Albert sent, albeit irregularly.

In 1925, Albert wrote in his will that the Nobel Prize money was his sons' inheritance. Mileva strongly objected, stating this money was hers and considered revealing her contributions to his work. Radmila Milentijevi\'{c} quote from a letter Albert sent her on 24 October 1925 (AEA 75-364). {\it "But you really gave me a good laugh when you started threatening me with your memories. Doesn't it ever dawn upon you for even a single second that no one would pay the least attention at all to your rubbish if the man with whom you are dealing had not perchance accomplished something important? When a person is a nonentity, there's nothing more to be said, but one should then be modest and shut up. I advise you to do so."}$^{[1]}$

Mileva remained silent but her friend Milana Bota told the Serbian newspaper Vreme in 1929 that they should talk to Mileva to find out about the genesis of special relativity, since she was directly involved, that Mileva had confided to her having {\it "participated to its creation"}.$^{[1]}$ On 13 June 1929, Mileva wrote to Helene Savi\'c: {\it "Milana could not help confiding our stories to the newspaper reporter, and I thought then that the matter was finished, so I did not talk about it at all. Such releases are not my style at all, but I believe it gave pleasure to Milana, and she probably thought that it would give me pleasure, too, and in a way would help me acquire certain rights with respect to Einstein in people's eyes. So she wrote to me, and so that's how we are going to take it, for otherwise the whole thing would be pointless."}$^{[4]}$

According to Krsti\'c $^{[2]}$, Mileva also spoke of her contributions to her mother and sister. She also wrote to her godparents explaining how she had always collaborated with Albert and how he had ruined her life, but asked them to destroy the letter. Her son, Hans Albert, told Krsti\'{c} how his parents {\it "scientific collaboration continued into their marriage, and that he remembered seeing his parents work together in the evenings at the same table."}$^{[2]}$ Hans Albert and his first wife, Frieda, tried to publish the letters Mileva and Albert had exchanged in their youth as well as those Albert had sent to his sons but were blocked in court by the Einstein's Estate Executors, Helen Dukas and Otto Nathan in an attempt to preserve the "Einstein's myth". They prevented other publications, including one from Krsti\'c $^{[2]}$ on his early findings in 1974. Krsti\'{c} mentions that Nathan even "visited" Mileva's apartment after her death in 1948.

Their letters and the numerous testimonies show that Mileva Mari\'{c} and Albert Einstein collaborated closely from their school days up to 1914. Albert referred to it repeatedly in his letters, like when he wrote: {\it "our work on relative motion.}" Their union was based on love and mutual respect, which allowed them together to produce such uncommon work. She was the first person to recognise his talent. Without her, he would never have succeeded. She abandoned her own aspirations, happy to work with him and contribute to his success, feeling they were one, {\it ein Stein}. Once started, the process of signing their work under his unique name became impossible to reverse. She probably agreed to it since her own happiness depended on his success. Why did Mileva remain silent? Being reserved and self-effaced, she did not seek honours or public attention. And even when she considered doing so, Albert's reaction was so brutal that she did not pursue it any further. It will probably never be possible to disentangle their individual contributions, as is often the case in any close collaborations.


\begin{thebibliography}{99}
\bibitem{(1)}
 Radmila Milentijevi\'{c}: Mileva Mari\'{c} Einstein: Life with Albert Einstein, United World Press, 2015.
\bibitem{(2)} 
Dord Krsti\'{c}: Mileva and Albert Einstein: Their Love and Scientific Collaboration, Didakta, 2004.
\bibitem{(3)}
Desanka Trbuhovi\'c-Gjuri\'{c}: Mileva Mari\'{c} Einstein: In Albert Einstein's shadow. Published in Serbian, 1969, German, 1982, and French, 1991. The quotes here were translated from French to English by me.
\bibitem{(4)} 
Milan Popovi\'{c}: In Albert's Shadow, the Life and Letters of Mileva Mari\'{c}, Einstein's First Wife, The John Hopkins University Press, 2003. I also used the original letters in German to improve on the English translation.
\bibitem{(5)}
Renn and Schulmann, Albert Einstein / Mileva Mari\'{c}, The Love Letters, Princeton University Press, 1992.
\bibitem{(6)} 
Peter Michelmore, Einstein, Profile of the Man, Dodd, Mead and Company, 1962.
\bibitem{(7)} 
R.S. Shankland, Conversation with Albert Einstein, Am. J. of Physics, 1962.\\
{\footnotesize  https://archive.org/stream/EinsteinConversations/Einstein\_djvu.txt}
\bibitem{(8)} 
Collected Papers of Albert Einstein\\ { https://einsteinpapers.press.princeton.edu/}

\end{thebibliography}
\end{document}